# Monte Carlo calculations of cryogenic photodetector readout of scintillating GaAs for dark matter detection

Stephen E. Derenzo
Lawrence Berkeley National Laboratory

**ABSTRACT**

The recent discovery that GaAs(Si,B) is a bright cryogenic scintillator with no apparent afterglow offers new opportunities for detecting rare, low-energy, electronic excitations from interacting dark matter. This paper presents Monte Carlo calculations of the scintillation photon detection efficiencies of optical cavities using three current cryogenic photodetector technologies. In order of photon detection efficiency these are: (1) Ge/TES: germanium absorbers that convert interacting photons to athermal phonons that are readout by transition edge sensors, (2) KID: kinetic induction detectors that respond to the breaking of cooper pairs by a change in resonance frequency, and (3) SNSPD: superconducting nanowire single photon detectors, where a photon briefly transitions a thin wire from superconducting to normal. The detection efficiencies depend strongly on the *n*-type GaAs absolute absorption coefficient $K_A$, which is a part of the narrow beam absorption that has never been directly measured. However, the high cryogenic scintillation luminosity of GaAs(Si,B) sets an upper limit on $K_A$ of 0.03/cm. Using that value and properties published for Ge/TES, KID, and SNSPD photodetectors, this work calculates that those photodetectors attached to opposing faces of a 1 cm$^3$ cubic GaAs(Si,B) crystal in an optical cavity with gold mirrors would have scintillation photon detection efficiencies of 35%, 25%, and 8%, respectively. Larger values would be expected for lower values of $K_A$.

Keywords:
dark matter detection
optical cavity
scintillating GaAs
transition edge sensor (TES)
kinetic induction detector (KID)
superconducting nanowire single photon detector (SNSPD)

## 1. Introduction

Dark matter is a major component of the universe and is known for its ability to gravitationally affect the paths of stars, galaxies, and photons. Understanding its properties is one of the greatest challenges in physics. However, in spite of many experiments using a variety of targets, it has not been detected directly through its interaction with standard matter. This paper is intended to aid in the design of new dark matter detection experiments that use the recently discovered scintillator GaAs(Si,B) [1-3]. It compares polished and ground GaAs surface treatments, and three cryogenic photodetector technologies currently under development.

### 1.1 Scintillating GaAs(Si,B)

GaAs(Si,B) is a luminous cryogenic scintillator [1-3] that has four emission bands that peak at 850, 930, 1070, and 1335 nm [2, 4-11]. These wavelengths lie in the deep valley between the valence to conduction band absorption that rises sharply for decreasing wavelengths below 816 nm and the optical phonon absorption that rises above 3 μm [12, 13]. It is different from conventional inorganic scintillators that require the separate migration of ionization electrons and holes to form radiative centers. Instead, the silicon doping in *n*-type GaAs provides a built-in population of delocalized electrons at the bottom of the conduction band and only the migration of ionization holes from the valence band to boron acceptors is required to form radiative centers (see Fig. 1).

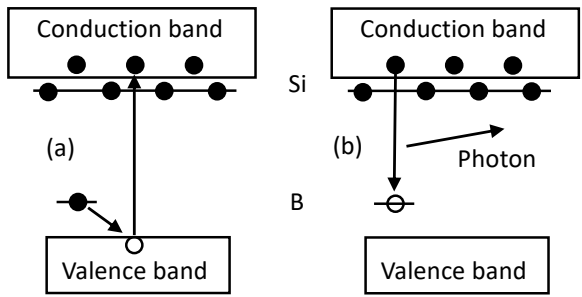

Figure 1. GaAs(Si,B) is doped with silicon donors to provide delocalized electrons in the conduction band and doped with boron acceptors above the valence band that are saturated with electrons. (a) When an ionization event produces a hole in the valence band, a boron ion loses an electron to fill that hole. (b) A conduction band electron combines with the boron hole to produce a photon.

GaAs(Si,B) is an attractive target for detecting rare, low-energy, electronic excitations from interacting dark matter for seven essential reasons, as follows:

1) Silicon donor electrons in GaAs have a binding energy that is among the lowest of all known *n*-type semiconductors. Free electrons above $8 \times 10^{15}$ per cm$^3$ are not "frozen out" and remain delocalized at cryogenic temperatures [14].
2) Boron and gallium are group III elements, so boron as an impurity primarily occupies the gallium site. However, a sufficient number occupy the arsenic site [15] to act as acceptors that efficiently trap ionization event holes from the valence band.
3) After trapping an ionization event hole from the valence band, the boron acceptors can combine radiatively with delocalized donor electrons to produce photons $\geq 0.2$ eV below the cryogenic band-gap energy (1.52 eV) (see Figure 1). This is an efficient radiative process that produces scintillation photons that are not absorbed by valence to conduction band transitions in the GaAs crystal [2, 3].
4) There is no detectable afterglow, as evidenced by the lack of thermally induced luminescence [1]. This is understandable because metastable radiative centers (i.e. trapped holes) are promptly filled by the delocalized *n*-type electrons [16].
5) GaAs has a high refractive index (~3.5) and the narrow-beam absorption coefficient at the emission wavelengths is proportional to the free electron density and typically several per cm [12, 13]. One would expect that almost all of the scintillation photons should be confined by total internal reflection and absorbed in the crystal, but this is not the case. Monte Carlo calculations that track scintillation photons in GaAs show that the high luminosity could be explained if most of the narrow beam absorption is not absolute absorption but a *novel* type of optical scattering from the conduction electrons with a cross section of about $5 \times 10^{-18}$ cm$^2$ that allows scintillation photons to escape total internal reflection [17]. This cross section is about $10^7$ times larger than Thomson scattering but comparable to the optical cross section of the conduction electrons in a metal mirror. Feynman path integral calculations of the interaction of optical photons with sheets of conduction electrons show how this optical scattering is possible [18]. If the sheet is much thinner than the wavelength, as in a metal mirror, reflection dominates. If the same conduction electrons are spread throughout an extended volume, as in *n*-type GaAs, scattering dominates.
6) *N*-type GaAs(Si,B) is commercially grown as 10 kg crystal ingots and sliced into thin wafers as substrates for electronic circuits. Boron oxide is used as an encapsulant to prevent the loss of arsenic during crystal growth, but also has the benefit of providing boron acceptors for scintillation.
7) GaAs(Si,B) scintillation photons can be detected with thermal absorbers coupled to transition edge sensors [19, 20], with microwave kinetic inductance detectors [21, 22], or with superconducting nanowire single photon detectors [23-25]. When used as pairs coupled to opposing faces, the time-correlated low backgrounds of these photodetectors and the apparent absence of afterglow from the GaAs(Si,B) could potentially have fewer "low energy excess" events below 200 eV that limits many dark matter detection experiments [20, 26-30].

**1.2 Photon absorber and transition edge sensor**

For over 20 years experiments have attempted to detect heat pulses from interacting dark matter. The CDMS (cryogenic dark matter search) experiments use germanium targets [31-33] and the CRESST (Cryogenic Rare Event Search with Superconducting Thermometers) experiments mainly use CaWO4 targets [28, 34, 35]. In both cases a superconducting element is operated at a stable temperature point in the transition between the normal and superconducting phase. When a heat pulse is received, the element becomes more resistive and a feedback circuit acts to keep the wire at the same temperature point. The transition edge is so sharp that μK changes in temperature can be recorded by the feedback circuit.

More recently this technology has been improved for the detection of individual photons absorbed in thin silicon wafers. Ref [19] reports an energy resolution of 3.86 eV rms using a 45.6 $cm^2$ x 1 mm silicon wafer. The CRESST collaboration reports the detection of individual 7.6 eV photons from a sapphire crystal using a 4 $cm^2$ x 0.4 mm silicon wafer [20]. The efficient detection of ~1 eV photons from GaAs(Si,B) will require an absorber with a lower band gap (such as germanium) , and a significant improvement in energy resolution.

**1.3 Kinetic inductance photo detectors**

Kinetic inductance detectors contain a superconducting wire whose inductance depends inversely on the density of charge carriers (Cooper pairs). It is used in an LC circuit with a resonance frequency in the GHz range. The absorption of a 0.3 eV photon breaks enough Cooper pairs to be detected by a change in the resonance frequency by one part in $10^6$ [22, 36-39]. Large area KID photodetectors have been proposed [40], including one for reading out scintillating GaAs [21].

**1.4 Superconducting nanowire and microwire single photon detectors**

SNSPDs contain a superconducting wire that is operated close to its critical current density. The energy deposited by a sub-eV photon is sufficient to transition a small section of the wire to the normal resistive state. The energy stored in the circulating current is dissipated in that section, and the rapid (ps) decrease in current is detected electronically. The superconducting current recovers as the wire cools. Small area SNSPDs are available commercially for coupling to optical fibers. Recent discoveries show that the wires can be made > 1 μm in width by optical lithography and cover larger areas [23, 41].

SNSPDs and KIDs have been proposed for dark matter detection experiments but the target masses are on the order of ng [22, 42, 43]. Using these as photodetectors for scintillating GaAs would increase the target mass by a factor of $10^9$, but would limit detection to electronic excitations above the 1.52 eV band gap.

This paper is organized as follows: Section 2 describes the Monte Carlo processes for tracking scintillation photons until they are absorbed by the GaAs, the mirrors, or the photodetectors. Section 3 presents results for the three cryogenic photodetectors. Section 4 discusses the implementation of optical cavities for dark matter detection. Section 5 presents conclusions. The Appendix describes the Monte Carlo calculations in sufficient detail for implementation in any scientific computer language.

## 2. Monte Carlo Processes

The Monte Carlo code tracks photons generated at random positions in a 1 cm³ cubic GaAs crystal and with isotropically random direction cosines until they are absolutely absorbed in the GaAs, by the six enclosing gold mirrors, or by the two detectors that are gap-coupled to opposing GaAs faces (Fig. 2). It is not assumed that the photodetectors can be built directly onto the GaAs crystal and a small gap is used between them in the calculations. Because of the high refractive index of GaAs, scintillation photons are internally and externally reflected many times before they can reach a photodetector. As a result, information about the point of origin is lost and all photons have the same detection probability, whether they are the result of a multi-photon emission or a single photon emission.

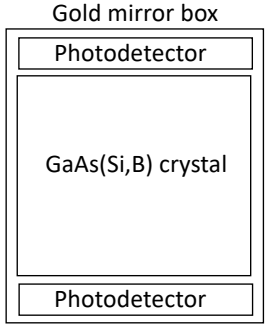

Figure 2. Optical cavity: a six-sided box of gold mirrors containing a scintillating GaAs crystal with photodetectors gap-coupled to two opposing faces.

In measurements of 29 GaAs samples with a wide range of silicon and boron doping levels, the luminosities in the different emission bands varied from sample to sample [3]. For the 12 samples with luminosities > 50 photons/keV the average emission wavelength was about 1100 nm. Using the data-based formula from ref. [44], the GaAs refractive index at 1100 nm and 301K is 3.465 ± 0.004. The refractive index decreases by 0.04 on cooling to 4K [45]. In this paper the value of 3.42 was used.

The following paragraphs summarize the calculations needed to track the scintillation photons as they are scattered, reflected, refracted, and absolutely absorbed. See the Appendix for a list of program steps.

When a photon encounters a polished GaAs surface from the inside or outside, the probabilities for reflection and refraction are given by the Fresnel equation in Appendix Note 2. In the case of reflection, the sign of the direction cosine perpendicular to the surface is reversed. In the case of refraction, the direction cosines are changed as described in Appendix Note 4.

When a photon encounters a Lambertian GaAs surface from the inside it reflects or refracts from a randomly oriented microfacet. Averaged over all angles the probabilities for reflection back into or escape out of the GaAs crystal are 0.9702 and 0.0298, respectively. For hemispherically random photons incident on a Lambertian GaAs surface from the outside, a separate Monte Carlo calculation found that the average probabilities for reflection or refraction are 0.3734 and 0.6266, respectively. In both cases the new direction cosines are hemispherically random.

When a photon encounters a gold mirror the probabilities for reflection and absorption are 0.98 and 0.02, respectively. Gold mirrors were chosen because their reflectance is high over all angles and over the different scintillation wavelengths. Since the GaAs crystal emits photons from its entire surface and in all directions isotropically, efficient constructive interference does not appear to be possible. This is in contrast to high detection efficiency optical stacks that work well for a narrow range of incident angles and wavelengths [46].

When a photon encounters a photodetector it can be transmitted, reflected, or absorbed. See Table 1 for the properties of the three cryogenic photodetectors used in this paper and Table 2 for a glossary of terms and parameters used. Photons are only detected if they are not reflected from nor penetrate through the detector. The optical penetration depth $D_P$ and reflectance $D_R$ are given by

$$D_P = \frac{\lambda}{4\pi k} \quad \text{and} \quad D_R = \frac{(n-1)^2 + k^2}{(n+1)^2 + k^2}$$

where $n$ is the real part and $k$ is the imaginary part of the refractive index of the detector material. A photon with direction cosine $d$ relative to the detector surface normal has a detection probability $Q_D$ given by:

$$Q_D = D_F(1 - D_R)\left(1 - \exp\left(\frac{-D_T}{d\, D_P}\right)\right)$$

where $D_F$ is the fill factor, $D_R$ is the reflectance, $D_T$ is the thickness, and $D_P$ is the optical penetration depth.

While the Monte Carlo calculation treats each photon individually, a useful figure of merit for a photodetector is the detection efficiency $Q_A$ for incident photons averaged over all angles of incidence. For isotropically random incidence, the average value of the cosine is 0.5 and the average path length in the detector material is twice its thickness.

$$Q_A = D_F(1 - D_R)\left(1 - \exp\left(\frac{-2D_T}{D_P}\right)\right)$$

Table 1 Constants for 1100 nm photons used in the calculations

| Photodetector technology | Ge/TES[19] | KID[22] | SNSPD[23] |
|---|---|---|---|
| Photon detector material | Ge | TiN | WSi |
| Detector active area fraction ($D_F$) | 1.0 | 0.85 | 0.4 |
| $n$ (real part of the refractive index) | 4.4 | 2.40 | 3.83 |
| $k$ (imaginary part of refractive index) | 0.11 | 4.62 | 2.88 |
| Detector reflectance ($D_R$) | 0.40 | 0.71 | 0.52 |
| Detector thickness $D_T$ (nm) | $10^6$ | 22 | 3.5 |
| Penetration depth $D_P$ (nm) | 800 | 19 | 30 |
| Isotropic detection probability $Q_A$ | 0.60 | 0.22 | 0.040 |

Table 2 - Glossary of terms used. The single photon detection efficiency $F_D$ is in boldface.

| | |
|---|---|
| SNSPD/WSi | Superconducting nanowire single photon detector with WSi wires |
| KID/TiN | Kinetic inductance detector with TiN wires |
| Ge/TES | Germanium photon absorber/athermal phonon transition edge sensor |
| Lambertian surface | Scatters light from any direction into a uniformly random hemisphere |
| $K_N$ | GaAs narrow-beam absorption coefficient (cm$^{-1}$) ($K_N = K_A + K_S$) |
| $K_A$ | GaAs absolute absorption coefficient (cm$^{-1}$) |
| $K_S$ | GaAs scattering coefficient (cm$^{-1}$) |
| $D_F$ | Detector fill factor (active area fraction) |
| $D_R$ | Detector active area reflectance |
| $D_T$ | Detector thickness (average path for isotropic photons = $2D_T$) |
| $D_P$ | Detector optical penetration depth (nm) |
| $F_e$ | Fraction of primary scintillation photons that escape the GaAs crystal |
| $P_e$ | Average photon path for primary photons escaping the GaAs crystal (cm) |
| **$F_D$** | **Fraction of scintillation photons absorbed by either detector** |
| $F_M$ | Fraction of scintillation photons absorbed by a gold mirror |
| $F_G$ | Fraction of scintillation photons absorbed in the GaAs crystal |
| $P_A$ | Average photon path at absorption (cm) |
| $M_R$ | Reflectance of enclosing gold mirror box |
| $Q_A$ | Average detector quantum efficiency for isotropic photons |

In this paper only time correlated photodetector readout of opposing faces is considered because of its overwhelming advantage in lower background rates over single face readout. For example, if each opposing detector has a dark count rate of $10^{-2}$/s, an accidental coincidence will occur in a $10^{-5}$ s timing window on the average once every 32 years. This random background is much lower than from other sources encountered in dark matter detection experiments.

## 3. Results

This section presents results for photons generated with random positions and random isotropic directions in a 1 cm$^3$ GaAs cube, and with photodetectors on two opposing faces (see Fig. 1). It explores a range of absolute absorption and scattering coefficients that have never been separately measured. However, their sum was measured by narrow beam experiments and was found to be proportional to the conduction electron concentration over a wide range [12, 13].

Figure 3 shows that for $K_A = 0.03$/cm (an upper limit consistent with the high scintillation luminosity[17]) the following conclusions can be made: (1) for polished faces the detection efficiency is low for low values of $K_S$ (solid curves) because a large fraction of the photons are trapped by total internal reflection, (2) for polished faces the detection efficiency increases to asymptotic limits as $K_S$ is increased, and (3) for Lambertian faces the detection efficiency is independent of $K_S$ because the rough surfaces provide a sufficient randomization of directions. Figure 4 shows that all the detection efficiencies increase for a decreased value $K_A = 0.003$/cm. The shapes of the curves are similar to those in Fig 3. Figure 5 shows the detection efficiency as a function of $K_A$ for $K_S = 1$/cm. Because of the long path lengths in the GaAs the efficiency decreases rapidly with increasing $K_A$.

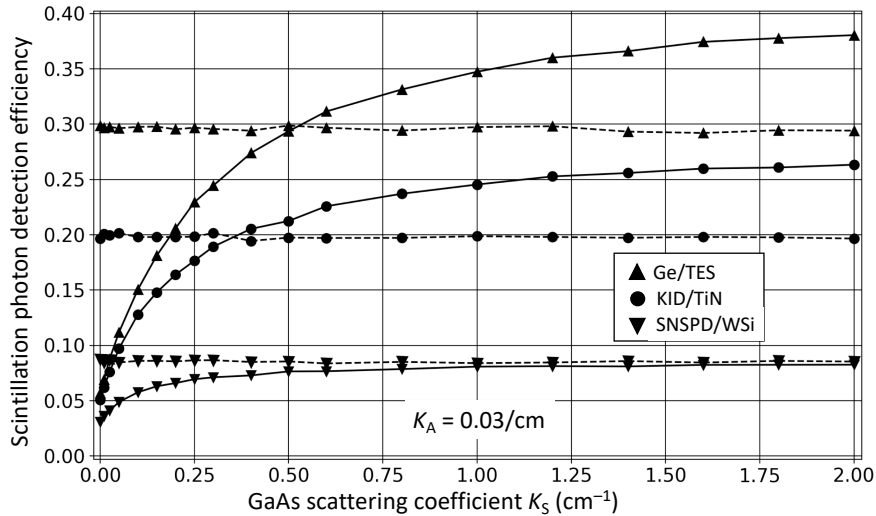

Figure 3. Optical cavity scintillation photon detection efficiency as a function of $K_S$ for $K_A = 0.03$/cm. Solid lines polished surfaces; dashed lines Lambertian surfaces.

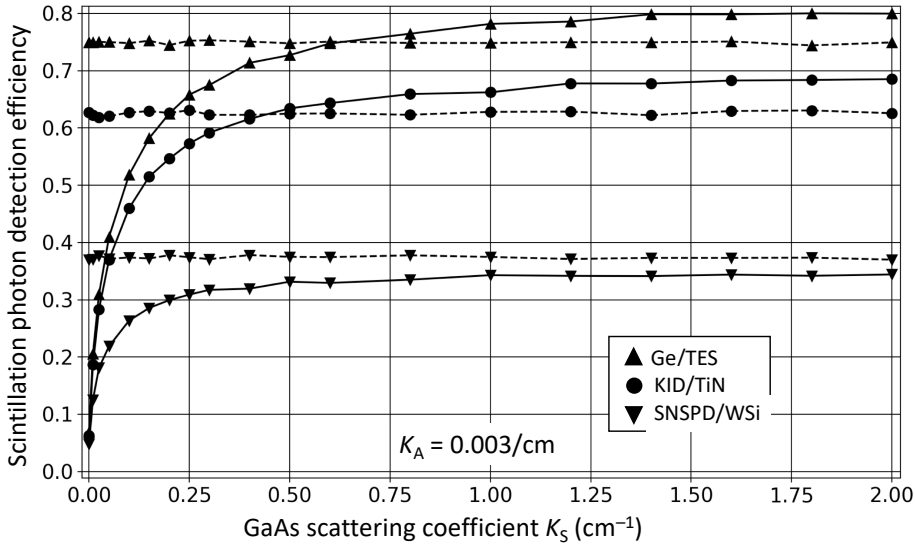

Figure 4. Optical cavity scintillation photon detection efficiency as a function of $K_S$ for $K_A = 0.003$/cm. Solid lines polished surfaces; dashed lines Lambertian surfaces.

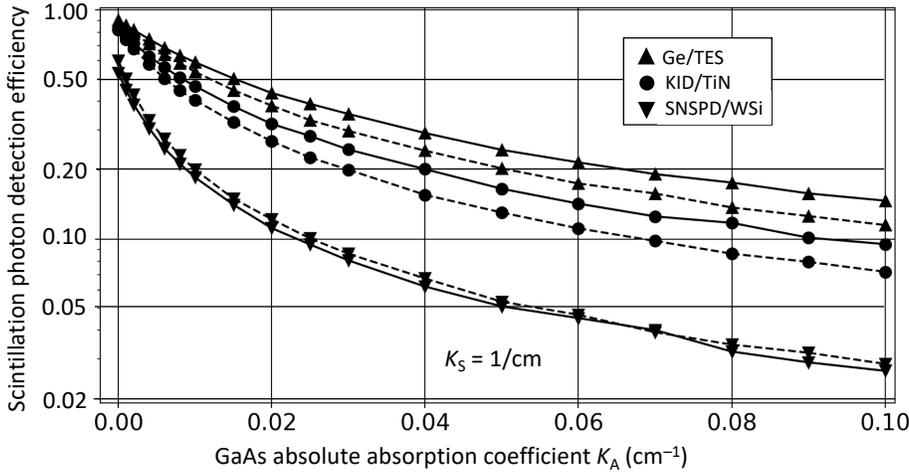

Figure 5. Optical cavity scintillation photon detection efficiency as a function of $K_A$ for $K_S = 1$/cm. Solid lines polished faces, dashed lines Lambertian faces.

Tables 3, 4, and 5 list results for $10^5$ photons generated in a 1 cm³ GaAs crystal. The rms statistical uncertainty for the fractions $F = F_e$, $F_D$, $F_M$, or $F_G$ is given by $\sqrt{F/10^5}$. For example, the rms uncertainties for $F = 0.1$ and 1 are 0.001 and 0.003, respectively.

Table 3 lines 1-6 lists the results for polished (P) faces and Ge/TES photodetectors on two opposing faces. For $K_S = K_A = 0$ (row 1), 94.7% of the photons are trapped by total internal reflection and 4.5% are absorbed by the detectors. Lines 2-6 show that for a constant narrow beam absorption $K_N = K_S + K_A = 1$/cm, when $K_A$ is decreased from 1/cm to 0/cm, (1) the primary scintillation photons have a larger probability $F_e$ of escaping the crystal after many internal reflections, which results in a larger path $P_e$ in the crystal before escape and (2) the fraction $F_D$ absorbed in the detectors increases, in spite of the increase in absorption $F_M$ by the mirrors.

Table 3 lines 7-12 lists the results for Lambertian (L) faces and Ge/TES photodetectors on two opposing faces. For $K_S = K_A = 0$ (row 7), the Lambertian surfaces provide a scattering mechanism that allows 90.5% of the photons to be absorbed by the detectors. This is much larger than for polished faces (row 1). Rows 2-6 show that for a constant narrow beam absorption $K_N = K_S + K_A = 1$/cm, when $K_A$ is decreased from 1/cm to 0/cm the effect on $F_e$, $P_e$, $F_D$, and $F_M$ is similar to the case of the polished faces.

The primary difference between polished and Lambertian faces is at low $K_S$, where trapping by total internal reflection dominates for polished faces.

Table 4 shows that the detection efficiency for the KID/TiN photodetector is slightly lower than that of the Ge/TES photodetector (Table 3), largely because the thickness is comparable to the optical penetration depth and it is more reflective (see Table 1). The values of the other parameters are also similar.

Table 5. shows that the detection efficiency of the SNSPD photodetector optical cavity is much lower than the other two because it is much thinner than the optical penetration depth and has a lower fill factor. The wires are thin so that the heat produced by a single photon can make part of the wire a normal conductor and increase the superconducting current density above the critical limit. By comparison, the KID is based on the small change in inductance that occurs when a photon brakes Cooper pairs and can be much thicker, and the Ge/TES is sufficiently thick so that virtually all photons are either reflected or absorbed.

Table 3. Optical cavity results for opposing Ge/TES photodetectors, for different scattering and absolute absorption coefficients, and for polished (P) and Lambertian (L) GaAs surfaces (see Table 2 for definitions of terms). The single photon detection efficiency $F_D$ is in boldface.

| Row | faces | $K_S$ | $K_A$ | $F_e$ | $P_e$ | $F_D$ | $F_M$ | $F_G$ | $F_T$ | $P_A$ |
|---|---|---|---|---|---|---|---|---|---|---|
| 1 | P | 0.00 | 0.00 | 0.132 | 6.08 | **0.0451** | 0.0082 | 0.0000 | 0.9467 | 6.44 |
| 2 | P | 0.00 | 1.00 | 0.065 | 0.94 | **0.0166** | 0.0020 | 0.9815 | 0.0000 | 0.99 |
| 3 | P | 0.50 | 0.50 | 0.119 | 1.76 | **0.0324** | 0.0030 | 0.9646 | 0.0000 | 1.94 |
| 4 | P | 0.90 | 0.10 | 0.402 | 5.96 | **0.1416** | 0.0154 | 0.8431 | 0.0000 | 8.44 |
| 5 | P | 0.99 | 0.01 | 0.869 | 12.86 | **0.5886** | 0.0610 | 0.3504 | 0.0000 | 34.96 |
| 6 | P | 1.00 | 0.00 | 1.000 | 14.81 | **0.9048** | 0.0952 | 0.0000 | 0.0000 | 54.16 |
| 7 | L | 0.00 | 0.00 | 1.000 | 18.79 | **0.9048** | 0.0952 | 0.0000 | 0.0000 | 68.14 |
| 8 | L | 0.00 | 1.00 | 0.048 | 0.95 | **0.0125** | 0.0011 | 0.9864 | 0.0000 | 1.00 |
| 9 | L | 0.50 | 0.50 | 0.093 | 1.82 | **0.0250** | 0.0028 | 0.9722 | 0.0000 | 1.97 |
| 10 | L | 0.90 | 0.10 | 0.344 | 6.55 | **0.1150** | 0.0120 | 0.8731 | 0.0000 | 8.82 |
| 11 | L | 0.99 | 0.01 | 0.841 | 15.90 | **0.5366** | 0.0578 | 0.4057 | 0.0000 | 40.95 |
| 12 | L | 1.00 | 0.00 | 1.000 | 18.85 | **0.9042** | 0.0958 | 0.0000 | 0.0000 | 68.74 |

Table 4. Optical cavity results for opposing KID photodetectors, for different scattering and absolute absorption coefficients, and for polished (P) and Lambertian (L) GaAs surfaces (see Table 2 for definitions of terms). The single photon detection efficiency $F_D$ is in boldface.

| Row | faces | $K_S$ | $K_A$ | $F_e$ | $P_e$ | $F_D$ | $F_M$ | $F_G$ | $F_T$ | $P_A$ |
|---|---|---|---|---|---|---|---|---|---|---|
| 1 | P | 0.00 | 0.00 | 0.130 | 6.09 | **0.0308** | 0.0086 | 0.0000 | 0.9606 | 6.47 |
| 2 | P | 0.00 | 1.00 | 0.064 | 0.94 | **0.0104** | 0.0022 | 0.9874 | 0.0000 | 0.99 |
| 3 | P | 0.50 | 0.50 | 0.118 | 1.76 | **0.0207** | 0.0044 | 0.9749 | 0.0000 | 1.95 |
| 4 | P | 0.90 | 0.10 | 0.402 | 5.99 | **0.0920** | 0.0189 | 0.8891 | 0.0000 | 8.93 |
| 5 | P | 0.99 | 0.01 | 0.870 | 12.97 | **0.4619** | 0.0945 | 0.4436 | 0.0000 | 44.76 |
| 6 | P | 1.00 | 0.00 | 1.000 | 14.78 | **0.8285** | 0.1715 | 0.0000 | 0.0000 | 79.93 |
| 7 | L | 0.00 | 0.00 | 1.000 | 18.75 | **0.8204** | 0.1796 | 0.0000 | 0.0000 | 104.91 |
| 8 | L | 0.00 | 1.00 | 0.048 | 0.95 | **0.0075** | 0.0016 | 0.9909 | 0.0000 | 1.00 |
| 9 | L | 0.50 | 0.50 | 0.094 | 1.81 | **0.0155** | 0.0032 | 0.9813 | 0.0000 | 1.98 |
| 10 | L | 0.90 | 0.10 | 0.345 | 6.57 | **0.0705** | 0.0154 | 0.9141 | 0.0000 | 9.28 |
| 11 | L | 0.99 | 0.01 | 0.840 | 15.90 | **0.4021** | 0.0881 | 0.5098 | 0.0000 | 51.85 |
| 12 | L | 1.00 | 0.00 | 1.000 | 18.92 | **0.8184** | 0.1816 | 0.0000 | 0.0000 | 105.79 |

Table 5. Optical cavity results for opposing SNSPD photodetectors, for different scattering and absolute absorption coefficients, and for polished (P) and Lambertian (L) GaAs surfaces (see Table 2 for definitions of terms). The single photon detection efficiency $F_D$ is in boldface.

| Row | | $K_S$ | $K_A$ | $F_e$ | $P_e$ | $F_D$ | $F_M$ | $F_G$ | $F_T$ | $P_A$ |
|---|---|---|---|---|---|---|---|---|---|---|
| 1 | P | 0.00 | 0.00 | 0.130 | 6.09 | **0.0070** | 0.0099 | 0.0000 | 0.9831 | 6.48 |
| 2 | P | 0.00 | 1.00 | 0.063 | 0.94 | **0.0028** | 0.0027 | 0.9945 | 0.0000 | 1.00 |
| 3 | P | 0.50 | 0.50 | 0.119 | 1.77 | **0.0059** | 0.0049 | 0.9893 | 0.0000 | 1.99 |
| 4 | P | 0.90 | 0.10 | 0.399 | 5.98 | **0.0262** | 0.0236 | 0.9502 | 0.0000 | 9.51 |
| 5 | P | 0.99 | 0.01 | 0.870 | 12.83 | **0.1839** | 0.1626 | 0.6535 | 0.0000 | 65.49 |
| 6 | P | 1.00 | 0.00 | 1.000 | 14.84 | **0.5298** | 0.4702 | 0.0000 | 0.0000 | 187.42 |
| 7 | L | 0.00 | 0.00 | 1.000 | 18.7 | **0.6006** | 0.3994 | 0.0000 | 0.0000 | 203.11 |
| 8 | L | 0.00 | 1.00 | 0.049 | 0.95 | **0.0029** | 0.0019 | 0.9952 | 0.0000 | 1.01 |
| 9 | L | 0.50 | 0.50 | 0.093 | 1.82 | **0.0061** | 0.0036 | 0.9904 | 0.0000 | 2.01 |
| 10 | L | 0.90 | 0.10 | 0.343 | 6.52 | **0.0276** | 0.0191 | 0.9533 | 0.0000 | 9.63 |
| 11 | L | 0.99 | 0.01 | 0.838 | 15.93 | **0.1965** | 0.1294 | 0.6741 | 0.0000 | 67.80 |
| 12 | L | 1.00 | 0.00 | 1.000 | 18.91 | **0.6043** | 0.3957 | 0.0000 | 0.0000 | 206.30 |

Figure 6 relates the single-photon detection efficiency in Figs 3-5 and Tables 3-5 with the probability that both detectors will detect time-correlated photons for 2, 5, and 10 simultaneously generated GaAs scintillation photons. If the optical cavity has a 100% detection efficiency, two simultaneous photons will be detected one in each detector (i.e. in time correlation) 50% of the time. While time-correlated detection reduces the detector pair event rate, the background rate is reduced by a much larger factor, resulting in a significant improvement in the statistical signal to noise ratio. For a single detector random background rate $10^{-2}$/s, detector pair time-correlation reduces the random background by a factor of $10^7$ for a $10^{-5}$ s time window.

Based on Fig. 2 of ref [16], dark matter particles with 5 MeV/c² mass and an electron scattering cross section of $10^{-36}$ cm² should produce on the order of 100 two-photon events in a 1 cm³ GaAs scintillator per yr. Using Figs. 4 and 5 with $K_A = 0.03$/cm, about 10% of those will be detected as time correlated.

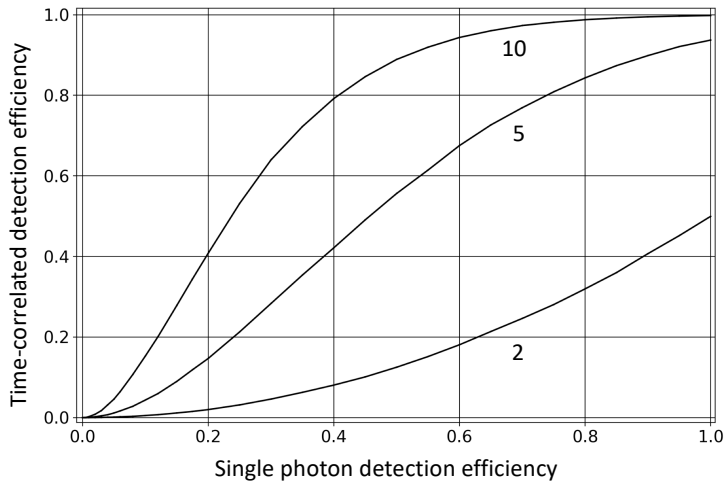

Figure 6. Time-correlated detection efficiency of opposing face detectors vs. the single photon detection efficiency for 2, 5, and 10 simultaneous GaAs scintillation photons.

In summary,
- Scintillation photons can only exit the GaAs crystal if their direction is in the exit cone of a polished face or of a rough face microfacet.

- Rough or polished, the scintillation photons must interact with the internal GaAs surfaces many times before they find an exit cone. As a result, their path lengths are much larger than the size of the GaAs crystal.
- Photons that exit the GaAs crystal must also reflect many times on the enclosing mirrors and pass through the GaAs many times before they can be absorbed by the detectors.
- Because of the long path lengths through the GaAs, a high detection efficiency requires a low GaAs absolute absorption. The detection efficiency decreases dramatically for absolute absorption coefficients $K_A$ above 0.03/cm.
- For polished faces, commercially available *n*-type GaAs has a sufficient concentration of *n*-type conduction electrons to provide a high detection efficiency for low $K_A$.
- Rough faces provide a high detection efficiency for low $K_A$ so optical polishing is optional.

## 4. Discussion

As seen in the figure and tables, the fraction of scintillation photons absorbed by the detectors depends critically on the GaAs absolute absorption coefficient ($K_A$) and much less on the scattering coefficient ($K_S$) if it is above 1/cm. Unfortunately, narrow beam measurements of the optical absorption of *n*-type GaAs do not make a distinction between absolute absorption and scattering [12, 13]. Rows 2 and 8 of Table 3 show that under the assumption that the narrow beam absorption is entirely absolute absorption ($K_S = 0$ and $K_A = 1$/cm), the detection efficiencies would be less than 0.02, inconsistent with the experimental evidence that cryogenic GaAs(Si,B) is a bright scintillator. New experiments are needed to determine the missing optical properties of cryogenic *n*-type GaAs. Failing that, experimental measurements of the detection efficiency of optical cavities of the type described here can provide estimates of $K_A$.

Polished and Lambertian GaAs surfaces are extremes of surface roughness. A realistic rough-ground surface is somewhere in between [47]. Since the detection efficiencies for polished and Lambertian surfaces in Figs. 2-4 and Tables 1-3 are reasonably close, it is safe to interpolate between them, according to the actual surface roughness.

The potential advantages of a Ge/TES photodetector are a large surface area and a high absorption efficiency. However, current energy resolution limits detection to pulses of many GaAs(Si.B) scintillation photons. The advantages of the KID photodetector is that it has the sensitivity to detect single sub-eV photons and the dynamic range to measure the energy deposited by a pulse of many photons without saturation. The SNSPD technology can also detect single sub-eV photons but the thickness of the wires is limited by the need to operate near the critical current density [48] and its absorption efficiency is lowest.

## 5. Conclusions

- Scintillating GaAs read out with current superconducting photodetectors offers new opportunities for detecting rare eV-level electronic excitations from interacting dark matter.
- The performance of optical cavities that use these technologies requires a better understanding of the scattering and absolute absorption coefficients ($K_S$ and $K_A$) of cryogenic *n*-type GaAs.
- For the upper limit $K_A = 0.03$/cm set by the scintillation luminosity, the detection efficiencies for the Ge/TES, KID, and SNSPD optical cavities with 1 cm$^3$ cubic GaAs scintillators are 35%, 25%, and 8%, respectively.
- For $K_A = 0$, the detection efficiencies for the Ge/TES, KID, and SNSPD optical cavities with 1 cm$^3$ cubic GaAs scintillators are 90%, 82%, and 60%, respectively.

**Appendix** - Detailed outline of the Monte Carlo computer code

The steps in the Monte Carlo calculation are described here in sufficient detail for implementation in any scientific programming language:

1. Define the following:
   Spatial coordinates and reflectance of the six bounding mirrors.
   Spatial coordinates of the six faces of the GaAs scintillator and its refractive index.
   Spatial coordinates, fill factor, reflectance, optical penetration depth, and thickness of the photodetectors.
   The absolute absorption coefficient $K_A$ and scatter coefficient $K_S$ in the GaAs.
2. Generate a new scintillation photon:
   Select random Cartesian coordinates $x, y, z$ from a uniform spatial distribution within the GaAs.
   Select a unit vector with random isotropic direction cosines according to Note 1.
   Go to 3. to start tracking the new photon.
3. At this point the photon has a specific spatial position and direction cosines.
   Compute the forward path lengths to the six mirror planes, to the six GaAs face planes and to both detector planes. Among those, find the shortest path length $P_1$.
   Go to 4.1 or 4.2, depending on the location of path $P_1$.
4.1 If path $P_1$ is in the GaAs volume, select random numbers $R_1$ and $R_2$ from a uniform distribution from 0 to 1. Compute a random absolute absorption path length $P_A = -\ln(R_1)/K_A$ and a random scatter path length $P_S = -\ln(R_2)/K_S$. Go to 4.1.1, 4.1.2, or 4.1.3, depending on the relative values of $P_A$, $P_S$, and $P_1$.
4.1.1 If $P_A$ is shorter than $P_S$ and $P_1$, tally the photon as absolutely absorbed in the GaAs and go to 2. to track a new photon.
4.1.2 If $P_S$ is shorter than $P_A$ and $P_1$, tally the photon as scattered in the GaAs. Advance the photon position by $\Delta x = P_S \cos(\theta_x)$, $\Delta y = P_S \cos(\theta_y)$, $\Delta z = P_S \cos(\theta_z)$. Select three new random direction cosines from an isotropic angular distribution according to Note 1 and go to 3. to continue tracking.
4.1.3 If $P_1$ is shorter than $P_A$ and $P_S$, go to 5.
4.2 If path $P_1$ is not in the GaAs volume, go to 5.
5. Advance the photon to the endpoint of path $P_1$: $\Delta x = P_1 \cos(\theta_x)$, $\Delta y = P_1 \cos(\theta_y)$, $\Delta z = P_1 \cos(\theta_z)$.
   Go to 5.1… 5.6, depending on the path endpoint.
5.1 If path $P_1$ ends at a *polished GaAs face from the inside* ($n_1 = 3.42$, $n_2 = 1$), compute the Fresnel reflectance $F$ according to Note 2.
   Select a random number $R_3$ from a uniform distribution from 0 to 1.
   If $R_3 \leq F$, the photon reflects back into the GaAs. Change the sign of the direction cosine that is perpendicular to the face and go to 3. to continue tracking. If the photon has totally internally reflected from three orthogonal faces and there is no scattering (see Note 3), tally the photon as trapped and go to 2. to generate a new photon.
   If $R_3 > F$, the photon refracts out of the GaAs. Generate new direction cosines according to Note 4 and go to 3. to continue tracking.
5.2 If path $P_1$ ends at a *Lambertian GaAs face from the inside*, the Fresnel reflectance averaged over all angles is $F = 0.9702$.
   Select a random number $R_4$ from a uniform distribution from 0 to 1.
   If $R_4 \leq F$, the photon reflects back into the GaAs at a random angle. Choose new hemispherically random direction cosines according to Note 1 and go to 3. to continue tracking.
   If $R_4 > F$, the photon refracts out of the GaAs with a random angle. Choose new hemispherically random direction cosines according to Note 1 and go to 3. to continue tracking.
5.3 If path $P_1$ ends at a *polished GaAs face from the outside* ($n_1 = 1$, $n_2 = 3.42$), compute the Fresnel reflectance $F$ according to Note 2.

Select a random number $R_5$ from a uniform distribution from 0 to 1.
If $R_5 \leq F$, the photon reflects from the GaAs face. Change the sign of the direction cosine that is perpendicular to the face and go to 3. to continue tracking.
If $R_5 > F$, the photon refracts into the GaAs face. Generate new direction cosines according to Note 4 and go to 3. to continue tracking.

5.4 If path $P_1$ ends at a *Lambertian GaAs face from the outside*, the Fresnel reflectance averaged over all angles is $F = 0.3734$.
Select a random number $R_6$ from a uniform distribution from 0 to 1.
If $R_6 \leq F$, the photon reflects from the GaAs face with a random angle. Compute new hemispherically random direction cosines according to Note 1. Go to 3. to continue tracking.
If $R_6 > F$, the photon refracts into the GaAs face with a random angle. Choose new hemispherically random direction cosines according to Note 1 and go to 3. to continue tracking.

5.5 If path $P_1$ ends at a *gold mirror face*, select a random number $R_7$ from a uniform distribution from 0 to 1.
If $R_7 \leq M_R$, the photon is reflected. Reverse the sign of the direction cosine perpendicular to the mirror face and go to 3. to continue tracking.
If $R_7 > M_R$, tally the photon as absorbed and go to 2. to track a new photon.

5.6 If path $P_1$ ends at a *detector face*, select random numbers $R_8$, $R_9$ and $R_{10}$ from a uniform distribution from 0 to 1
If $R_8 > D_F$, the photon has passed through gaps in the active detector area. Go to 3. to continue tracking.
If $R_9 \leq \exp(-D_T/D_P)$, the photon penetrated the active detector area without interacting. Go to 3. to continue tracking.
If $R_9 > \exp(-D_T/D_P)$ and $R_{10} \leq D_R$, the photon reflects from the detector. Reverse the sign of the direction cosine perpendicular to the detector face and go to 3. to continue tracking.
If $R_9 > \exp(-D_T/D_P)$ and $R_{10} \geq D_R$, the photon is absorbed in the detector. Tally as detected and go to 2. to generate a new photon.

Note 1: To select a random unit vector from a uniformly isotropic angular distribution, select three random trial cosines $d_x$, $d_y$, and $d_z$, each from a uniform distribution between $-1$. and $+1$ (for a hemisphere replace the $-1$ or $+1$ with 0, depending on the surface normal). Choose the first set that satisfies the condition $D^2 = d_x^2 + d_y^2 + d_z^2 \leq 1$. Scale to a unit vector by dividing each trial cosine by D: $\cos(\theta_x) = d_x/D$; $\cos(\theta_y) = d_y/D$; $\cos(\theta_z) = d_z/D$.

Note 2: The reflectance of the interface from refractive index $n_1$ to $n_2$ at incidence angle $\theta$ is given by the Fresnel equation:

$$F(\theta, n_1, n_2) = 0.5 \left( \frac{n_1 \cos(\theta) - n_2 \sqrt{1 - \left(\frac{n_1}{n_2}\sin(\theta)\right)^2}}{n_1 \cos(\theta) + n_2 \sqrt{1 - \left(\frac{n_1}{n_2}\sin(\theta)\right)^2}} \right)^2 + 0.5 \left( \frac{n_1 \sqrt{1 - \left(\frac{n_1}{n_2}\sin(\theta)\right)^2} - n_2 \cos(\theta)}{n_1 \sqrt{1 - \left(\frac{n_1}{n_2}\sin(\theta)\right)^2} + n_2 \cos(\theta)} \right)^2$$

Note 3: For a polished GaAs crystal without scattering, a photon is trapped by total internal reflection when the absolute values of all three direction cosines are less than the critical cosine for total internal reflection. For internal photons ($n_1 > n_2$) the critical angle for total internal reflection is $\arcsin(n_2/n_1)$. At each reflection the direction cosine parallel to the face normal changes sign but the absolute value does not change.

Note 4: To refract a photon from $n_1$ to $n_2$ first use Snell's law to change the angle relative to the face normal: $n_1 \sin(\theta_1) = n_2 \sin(\theta_2)$. The new direction cosine parallel to the face normal is $\cos(\theta_2)$. Since the optical momentum in the other two directions is conserved, multiply those direction cosines by $n_1/n_2$. The result is a set of refracted unit vector direction cosines.

## Acknowledgements

I thank M. Garcia-Sciveres, J. Luksin, F. Moretti and M. Shaw for helpful discussions, and the Lawrence Berkeley National Laboratory for remote access to the literature.